\documentclass[aps,prb,twocolumn,showpacs,tightenlines]{revtex4}
\usepackage{graphics}

\begin{document}

\title{Coulomb Blockade Oscillations of Conductance 
\\
at Finite Energy Level Spacing in a Quantum Dot}

\author{Serguei Vorojtsov}
\affiliation{Department of Physics, Duke University, Durham, NC 27708-0305}

\date{\today}

\begin{abstract}

We find an analytical expression for the conductance 
of a single electron transistor in the regime when
temperature, level spacing, and charging energy
of a grain are all of the same order.
We consider the model of equidistant energy levels in a grain 
in the sequential tunneling approximation.
In the case of spinless electrons 
our theory describes transport through a dot 
in the quantum Hall regime. 
In the case of spin-$\frac{1}{2}$ electrons we 
analyze the line shape of a peak, 
shift in the position of the peak's maximum as a function of temperature, 
and the values of the conductance in the odd and even valleys.
\end{abstract}

\pacs{73.23.Hk, 73.23.-b, 73.43.Jn}

\maketitle

\section{Introduction}

\begin{figure}[b]
\resizebox{.42\textwidth}{!}{\includegraphics{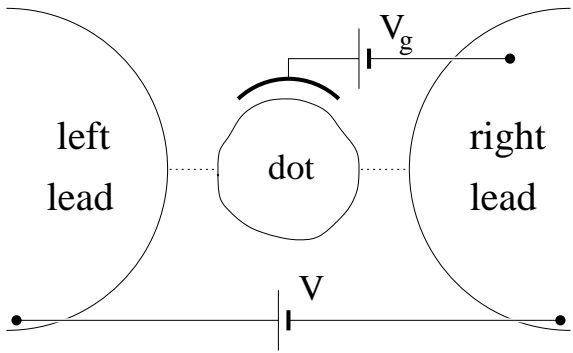}}
\caption{\label{Fig1}
Scheme of the Coulomb blockade setup.}
\end{figure}

Recent progress in mesoscopic fabrication techniques has
made possible not only the creation of more sophisticated 
devices but also greater control over their properties. 
Electron systems confined to small space regions, quantum dots,
and especially their transport properties have been studied
extensively for the last decade.\cite{ralph,fabr}
In particular, an individual ultra-small metallic grain of
radius less than $5$~$nm$ was attached to two leads via
oxide tunnel barriers, thus forming a single electron
transistor (SET).\cite{ralph}
Applying bias voltage, $V$, between two leads allows one
to study transport properties of the system, Fig.~\ref{Fig1}.
Alternatively, a SET can be formed by depleting two-dimensional
electron gas at the interface of GaAs/AlGaAs heterostructure
by applying negative voltages to the metallic surface gates.\cite{fabr}

In this paper we will assume that the bias voltage is infinitesimally
small, $V\to 0$. This corresponds to the linear response regime.
In order to tunnel onto the quantum dot, an electron in the left
lead has to overcome a charging energy, $E_C=e^2/2C$, where $e>0$ is 
the elementary charge; $C$ is the capacitance of the quantum dot.
If $T\ll E_C$ then conductance through the system is exponentially 
suppressed. This phenomenon is called the Coulomb blockade.
However if we apply a voltage, $V_g$, to the additional gate 
capacitively coupled to the dot, the Coulomb blockade can be lifted.
Indeed, changing $V_g$ one can shift the position of energy
minimum so that energies of the quantum dot with $N_e$ and $N_e+1$
electrons will become equal and an electron can freely jump 
from the left lead onto the dot and then jump out into the other lead. 
Thus, current event has occurred and a peak in the conductance, $G$,
corresponding to this gate voltage is observed. By changing the gate 
voltage one can observe an oscillation of the conductance or Coulomb 
blockade oscillations.

One-particle energy levels in the quantum dot, $\{E_i\}$, are given
by the solution of the Schr\"odinger equation in the quantum dot's
potential. The mean spacing between these energy levels is $\delta E$.
The conventional assumption that $E_C\gg\delta E$ 
is not valid in the case of sufficiently small dots.
In fact, in the recent experiments,\cite{porath,park}
where a $C_{60}$ molecule has acted as a quantum dot,
the level spacing is of order charging energy. Experiment\cite{porath}
was performed at $T=4.2 K$ as well as at room temperature.
In other experiments\cite{heiblum,bockrath} with quantum
dot formed by depleting 2DEG\cite{heiblum} and ropes of
carbon nanotubes acting as a quantum dot,\cite{bockrath}
charging energy is only three times larger than the spacing $\delta E$.

Though Coulomb blockade oscillations have been studied in a
number of important limiting cases,~\cite{kulik,glazman,matveev96,beenakker} 
the problem in the case when values of $E_C$, $\delta E$, and $T$ are 
all of the same order has not been theoretically addressed.
Let us note that energy levels of the quantum dots are random
and obey Wigner-Dyson statistics with the fluctuation of order
of their mean.\cite{mehta}
Nonetheless to go as far as possible in the analytical treatment 
of the problem we have to assume that energy levels in the quantum 
dot are equidistant. In this paper we derive an analytical expression 
for the linear conductance, $G=I/V|_{V\to 0}$, in the case of spinless 
as well as spin-$\frac{1}{2}$ fermions.

In Sec. II, we describe our model and the assumptions involved. 
We write the model assuming spin-$\frac{1}{2}$ fermions. 
In Sec. III, we consider the linear conductance in the case of spinless 
fermions. We obtain an analytical expression for the conductance and 
analyze its limiting cases. 
In Sec. IV, we consider one possible application of the Sec. III results, 
namely tunneling through the edge states in a quantum dot placed into 
a strong magnetic field. 
In Sec. V, the linear conductance as well as its properties in the case 
of spin-$\frac{1}{2}$ fermions is considered.
In Sec. VI, we summarize our findings.

\section{The model}

\begin{figure}[b]
 \resizebox{.42\textwidth}{!}{\includegraphics{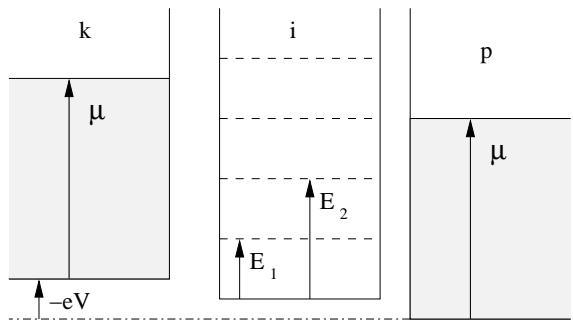}}
\caption{\label{Fig2}
Electrostatic potential energy along a line through the 
tunnel junctions.}
\end{figure}

Hamiltonian of the system in question is
\begin{equation}
{\hat H}={\hat H_{l}}+{\hat H_{d}}+{\hat T}.
\label{hamilt}
\end{equation}
Here, the first term is the Hamiltonian of noninteracting electrons 
in the left and right leads:
\begin{eqnarray}
{\hat H_{l}} = 
\sum_{k \sigma}E_{k}c_{k\sigma}^{\dagger}c_{k\sigma}+ 
\sum_{p \sigma}E_{p}c_{p\sigma}^{\dagger}c_{p\sigma},
\end{eqnarray}
where a continuum of states in each lead, 
$\left| k \sigma \right>$, $\left| p \sigma \right>$
is assumed;
$E_k$, $E_p$ and $c_{k\sigma}$, $c_{p\sigma}$ are the energies and 
electron annihilation operators in the left and right leads, 
respectively; $\sigma$ stands for the $z$-component of spin.
The chemical potentials of the leads, $\mu\gg E_C,\delta E,T$, 
are shifted according to the bias voltage, $V$, 
applied, Fig.~\ref{Fig2}. 
We will assume that leads are in thermal equilibrium at temperature 
$T$ and, thus, occupied according to the Fermi-Dirac distribution.

The second term in~(\ref{hamilt}) is the Hamiltonian 
of the quantum dot:
\begin{eqnarray}
{\hat H_{d}}=\sum_{i\sigma}E_{i}
c_{i\sigma}^{\dagger}c_{i\sigma}
+{\hat U},
\end{eqnarray}
where first term is the kinetic energy of electrons in the quantum dot:
$\{E_{i}\}$ is a discrete set of the quantum dot's energy levels; 
$c_{i\sigma}$'s are the annihilation operators. The second term, ${\hat U}$ 
describes the electron-electron interaction in the quantum dot.
We adopt the simplest model for the interaction, namely, the constant 
interaction model. In this model the Coulomb interaction of the electrons 
depends only on the total number of electrons in the quantum dot:
\begin{eqnarray}
U({\hat N})=E_{C}{\hat N}^{2}-eV_{e}{\hat N},
\end{eqnarray}
where 
$N=\sum_{i\sigma}c_{i\sigma}^{\dagger}c_{i\sigma}-N_{i}$
is the total number of excess electrons;
$N_i$ is the total number of positively charged ions.
The second term is the contribution from external charges.
They are supplied by the ionized donors and the gate:
$V_{e} = V_{d} + a V_{g}$, where $a$ is a function of the 
capacitance matrix elements of the system. Thus, $V_{e}$ can 
be varied continuously by changing gate voltage, $V_{g}$.
$U(N)$ can be rewritten as 
\begin{eqnarray}
U(N)=E_C(N-N_g)^2+\mbox{Const},
\end{eqnarray}
where $N_g=eV_{e}/2E_C$ is the dimensionless gate voltage.

The third term in~(\ref{hamilt}) is the tunneling Hamiltonian:
\begin{eqnarray}
{\hat T}=
\sum_{ki\sigma}\left( t_{ki}{c}_{k\sigma}^{\dagger}
{c}_{i\sigma}+h.c.\right)+
\sum_{pi\sigma}\left( t_{pi}{c}_{p\sigma}^{\dagger}
{c}_{i\sigma}+h.c.\right),
\end{eqnarray}
where $t_{ki}$ and $t_{pi}$ are matrix elements of tunneling into
the left and right leads, respectively.

We assume that the dot is weakly coupled to the leads;
that is, the conductances of the dot-lead junctions
are small: $G^{l,r} \ll e^2/h$, where $h$ is Planck's constant.
Equivalently, the widths of the quantum dot's energy levels 
contributing to the conductance, $\Gamma_i=\Gamma_i^l+\Gamma_i^r$,
must be small compared to spacing between them: $\Gamma_i\ll\delta E$.
This, together with $\Gamma_i \ll T$ assumption, allows us to 
characterize the state of the dot by a set of occupation 
numbers, $\{ n_{i\sigma}\}$.~\cite{beenakker}

\section{Linear conductance in the spinless case}

The model formulated above has been studied by Beenakker 
in the sequential tunneling approximation;\cite{beenakker}
that is, conservation of energy was assumed in each tunneling 
process, and cotunneling was neglected. Therefore, to find the 
stationary current, kinetic equation considerations can be applied.
In the linear response regime an analytical formula for the conductance 
has been obtained. In the case of spinless fermions:~\cite{beenakker}
\begin{eqnarray}
G &=& \frac{e^2}{h T} \sum_{i=1}^{\infty} 
\frac{\Gamma_i^l \Gamma_i^r}{\Gamma_i^l + \Gamma_i^r}
\sum_{N_e=1}^{\infty} P_{eq}(N_e) F_{eq}(E_i|N_e)
\nonumber \\
&& \times [1-n_F(E_i - \mu + U(N)- U(N-1)],
\label{spinlessg}
\end{eqnarray}
where
$$
\Gamma_i^l= 2\pi \sum_k |t_{ki}|^{2} 
\delta\left[ E_i-E_k+U(N)-U(N-1)\right]
$$
and
$$
\Gamma_i^r= 2\pi \sum_p |t_{pi}|^{2} 
\delta\left[ E_i-E_p+U(N)-U(N-1)\right]
$$
are widths of the quantum dot's level $i$ associated with
tunneling into the left and right leads, respectively;
$P_{eq}(N_e)$ is the equilibrium probability that the 
quantum dot contains $N_e$ electrons; 
$F_{eq}(E_i|N_e)$ is the occupation number of level $i$ 
given that the dot contains $N_e$ electrons; 
$n_F(E)$ is the Fermi-Dirac distribution; 
and $\mu$ is the chemical potential in the leads.

\begin{figure}[b]
\resizebox{.42\textwidth}{!}{\includegraphics{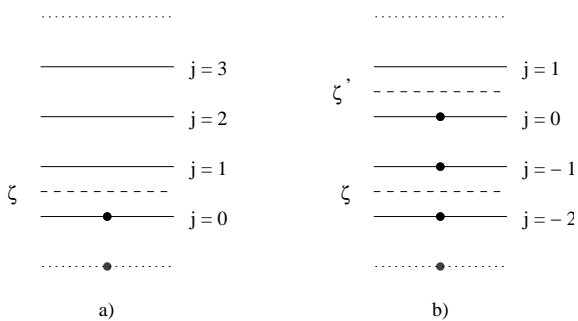}}
\caption{\label{Fig3}
a) occupation numbers in the canonical ensemble 
for the equidistant energy levels at $T=0$; 
b) mapping sum over $i$ onto sum over $j$ 
($N=2$ case is shown).}
\end{figure}

The quantity $F_{eq}(E_i|N_e)$ in~(\ref{spinlessg}) is the most 
non-trivial one to calculate. It is the occupation number of 
the level $i$ in the canonical ensemble ($N_e$ is fixed). 
In the limit $\delta E/T\to 0$, $F_{eq}(E_i|N_e)$ 
becomes a Fermi-Dirac distribution with the appropriately 
chosen chemical potential: $\tilde\mu=(E_{0}+E_{1})/2$,
where $E_{0}$ corresponds to the energy of the last 
occupied energy level at $T=0$, Fig.~\ref{Fig3}a; 
$E_{1}$ corresponds to the energy of the first empty energy 
level at $T=0$. In the opposite limit $\delta E/T\to\infty$, 
the Fermi-Dirac distribution with $\tilde\mu =(E_{0}+E_{1})/2$ 
apparently breaks down: the occupation number of level $j=1$, 
for example, see Fig.~\ref{Fig3}a, is $n_1=e^{-\delta E/T}$, 
not $e^{-\delta E/2T}$ as the Fermi-Dirac distribution would 
predict.\cite{beenakker}

The occupation number in question is~\cite{beenakker}
\begin{eqnarray}
F_{eq}(E_i|N_e)&=&\frac{1}{P_{eq}(N_e)}\sum_{\{ n_l\}} P_{eq}(\{ n_l\})
\delta_{n_i,1} \delta_{N_e,\sum n_l}
\nonumber \\
&=&e^{\beta F(N_e)} \sum_{\{ n_l\}}e^{-\beta \sum E_l n_l}
\delta_{n_i,1}\delta_{N_e,\sum n_l},
\label{tough}
\end{eqnarray}
where $P_{eq}(\{ n_l\})$ is the equilibrium probability of the
$\left| \{ n_l\}\right>$ state of the quantum dot; $\beta =1/T$;
and the detailed definition of $F(N_e)$ will follow. 
The reason for writing this equation is to show that analytical 
calculation of the occupation numbers is hardly possible for 
arbitrary quantum dot's energy level structure, $\{ E_i\}$. 

The only way to overcome this difficulty is 
to assume that energy levels in the quantum dot are equidistant.
Then one can use the bosonization technique\cite{haldane}
(see Appendix) 
to find the exact analytical expression 
for the occupation numbers in the canonical ensemble.
It was done by Denton, Muhlschlegel, and Scalapino:~\cite{denton}
\begin{equation}
F_{eq}(E_i | N_e) \equiv n_j = \sum_{m=1}^{\infty} (-1)^{m-1} 
e^{- \frac{1}{2}[m^2 + (2j-1)m] \frac{\delta E}{T}},
\label{dent}
\end{equation}
where
\begin{eqnarray}
j =\frac{E_i-\zeta}{\delta E}+\frac{1}{2}=\mbox{integer},
\label{nzero}
\end{eqnarray}
$\delta E/T\equiv\delta =\mbox{const}$, 
$\zeta$ is the energy corresponding to the
highest occupied energy level at $T=0$ plus
$\delta E/2$, Fig.~\ref{Fig3}a.
This quantity $\zeta$ 
is somewhat similar to the chemical potential of a dot, 
though, strictly speaking, 
the chemical potential is not well-defined for a dot 
in the canonical ensemble.
The difference, $\zeta -\mu$, is a linear function
of the gate voltage.
Therefore, by properly adjusting ``zero'' value 
of the gate voltage it can put to zero.
Hereinafter, we assume that $\zeta -\mu=0$.

To calculate the conductance we also need to know $P_{eq}(N_e)$, 
the probability that the dot, in thermodynamic equilibrium with 
the reservoirs, contains $N_e$ electrons. It can be calculated 
in the grand canonical ensemble:
\begin{equation}
P_{eq}(N_e) = \frac{1}{Z_{\mu}} e^{- \beta \varphi (N_e)},
\label{peq}
\end{equation}
where $Z_{\mu} = \sum_{N_e = 0}^{\infty} e^{- \beta \varphi (N_e)}$ 
is the grand  partition function; $\varphi (N_e)$ is the 
thermodynamic potential of the quantum dot. It can be expressed
via free energy of the dot's internal degrees of freedom, $F(N_e)$:
\begin{eqnarray}
\varphi (N_e) = F(N_e) - \mu N_e + U(N),
\end{eqnarray}
hence,
\begin{equation}
P_{eq}(N_e)=\frac{1}{Z_{\mu}}e^{- \beta U(N)}Z(N_e),
\label{peq2}
\end{equation}
where
$$
Z(N_e)=e^{-\beta [F(N_e)-\mu N_e]}
=\sum\limits_{\{ n_{i}\}}e^{-\beta\sum (E_i-\mu)n_{i}}\delta_{N_e,\sum n_{i}}
$$
is the partition function in the canonical ensemble. In the 
last expression the sum is taken over all possible states, 
$\left| \{ n_{i}\}\right>$ of the quantum dot.
To calculate $Z(N_e)$ explicitly we need to assume 
that energy levels of the dot are equidistant:
\begin{equation}
Z(N_e) = e^{-(N_e - N_{i})^2 \delta /2} Z_{exc},
\label{zne}
\end{equation}
where $N_e=N_{i}$ corresponds to the equilibrium number 
of the excess electrons $(\zeta =\mu)$; 
$Z_{exc}$ is the partition function of the thermal excitations.
Now, let us substitute Eq. (\ref{zne}) in Eq. (\ref{peq2}):
\begin{eqnarray}
P_{eq}(N_e) &=& \frac{1}{Z_{\mu}} e^{- \beta U(N)}
e^{-(N_e-N_{i})^2 \delta /2} Z_{exc} 
\nonumber \\
&=& \frac{1}{D'_{0}} e^{-\beta [U(N)+N^2\delta E/2]},
\label{pofne}
\end{eqnarray}
where
\begin{eqnarray}
D'_{0}=\sum_{N=-\infty}^{\infty}e^{-\beta [U(N)+N^2\delta E/2]}.
\end{eqnarray}
Here, we have extended one limit of the sum to infinity since 
the Fermi energy is the largest energy scale of the problem.

Thus, we are well-equipped to calculate the conductance 
in the case of equidistant energy levels in the quantum 
dot at arbitrary ratio $\delta$.
The widths of energy levels, $\Gamma_{i}^{l}$ and $\Gamma_{i}^{r}$,
in the quantum dot are energy dependent, random 
quantities. Let us assume that quantum dot is weakly coupled 
to the leads via multichannel tunnel junctions:
$G^{l,r}=G^{l,r}_{1}N_{ch}\ll e^2/h$,
where $N_{ch}$ is the number of channels;
$G^{l,r}_{1}$ is the conductance of one channel.
Experimentally, this situation corresponds to
the metallic grain coupled to the leads via oxide tunnel
barriers.\cite{ralph} This setup allows one to 
decrease fluctuations of the energy levels' widths,
$\Gamma_{i}^{l}$ and $\Gamma_{i}^{r}$, by a factor
of $\sqrt{N_{ch}}$.
We also assume that the widths are slowly changing
functions of the energy, $E_{i}$.
Then, $\Gamma_{i}^{l}$ and $\Gamma_{i}^{r}$ can be
considered constants and evaluated at the chemical potential:
$\Gamma_i^l \approx \Gamma_{\mu}^l \equiv \Gamma^l$; 
$\Gamma_i^r \approx \Gamma_{\mu}^r \equiv \Gamma^r$.
There is a simple relation between these widths and conductances
of the corresponding junctions. In the case of spinless fermions:
\begin{equation}
   \Gamma^l=\frac{h G^l}{e^2}\delta E,
~~~\Gamma^r=\frac{h G^r}{e^2}\delta E.
\label{glgr0}
\end{equation}
Let us substitute Eq.~(\ref{pofne}) in Eq.~(\ref{spinlessg}):
\begin{eqnarray}
G&=&\frac{G^lG^r}{G^l+G^r}\frac{\delta}{D'_{0}}
\sum_N e^{-\beta [U(N)+N^2\delta E/2]}
\sum_i F_{eq}(E_i|N_e)
\nonumber \\
&&\times \left\{ 1-n_F \left[ E_i-\mu +U(N)-U(N-1) \right] \right\}.
\label{interm}
\end{eqnarray}
To take advantage of the expression for occupation numbers, 
Eq.~(\ref{dent}), we need to map the sum over $i$ onto the sum over $j$. 
As illustrated in Fig.~\ref{Fig3}, the mapping rule depends 
on the total number of excess electrons, $N$ 
(compare with Eq.~(\ref{nzero}) written for $N=0$):
\begin{eqnarray}
F_{eq}(E_i|N_e) &=& n_j,\nonumber
\\
E_i-\mu =E_i-\zeta'+(\zeta'-\zeta)
&=&\left( j-\frac{1}{2}\right)\delta E+N\delta E,\nonumber
\end{eqnarray}
where $\zeta'=\zeta'(N)$ is the energy of the highest
occupied energy level in the dot with $N$ excess
electrons at $T=0$ plus $\delta E/2$.
We will also use the following identities:
\begin{eqnarray}
U(N)+\frac{\delta E}{2}N^2
=\left( E_C+\frac{\delta E}{2}\right) N^{2}-eV_{e}N
\nonumber \\
=\left( E_C+\frac{\delta E}{2}\right)
\left( N-\Delta_0 +\frac{1}{2}\right)^{2}+{\cal C}_{1},
\end{eqnarray}
where
${\cal C}_{1}=-\left(eV_{e}\right)^{2}/2(2E_C+\delta E)$;
\begin{eqnarray}
\Delta_0 \equiv \frac{eV_{e}}{2E_C+\delta E}+\frac{1}{2}
\end{eqnarray}
has been chosen so that $\Delta_0 =0$ corresponds to the 
maximum of the conductance peak; and
\begin{eqnarray}
U(N)-U(N-1)=(2N-1)E_C-eV_{e}
\nonumber \\
=(2N-1)E_C-(2E_C+\delta E)(\Delta_0-1/2).
\end{eqnarray}
Substituting these results in Eq.~(\ref{interm}), we obtain
\begin{equation}
\frac{G(\Delta_0)}{G_{\infty}}=
\frac{\delta}{D_0}
\sum_N e^{-\varepsilon_0\left( N-\Delta_0+\frac{1}{2}\right)^2}
\sum_j\frac{n_j}{e^{-j\delta -2(N-\Delta_0)\varepsilon_0}+1},
\label{g0}
\end{equation}
where
\begin{eqnarray}
D_0 = \exp \left(\beta{\cal C}_{1}\right) D'_0
=\sum_N e^{-\varepsilon_0\left( N-\Delta_0+\frac{1}{2}\right)^2};
\end{eqnarray}
$\varepsilon_0 \equiv \beta (E_C + \delta E/2)$; 
$G_{\infty} \equiv G^l G^r/(G^l + G^r)$ is the classical,
$E_C, \delta E \ll T \ll \mu$, limit of the conductance. We have 
also used the identity: 
$1-n_F(E) = (1+e^{-\beta E})^{-1}$. 
Formula (\ref{g0}) is the general expression for the linear 
conductance in the spinless case for equidistant energy levels 
in the quantum dot at arbitrary values of $E_C$, $\delta E$ and $T$. 

One can immediately prove the following properties of the 
conductance (\ref{g0}). 
First of all, $G(\Delta_0)=G(\Delta_0+M)$, 
where $M$ is an integer. In the gate voltage units, 
$\Delta V_e=(2E_C+\delta E)/e$ is a period of the conductance
oscillations. This property reflects symmetry with respect to adding
(removing) an electron to the quantum dot. 
Secondly, due to the electron-hole symmetry, conductance is an even 
function of $\Delta_0$: $G(\Delta_0)=G(-\Delta_0)$.
\begin{figure}[b]
\resizebox{.42\textwidth}{!}{\includegraphics{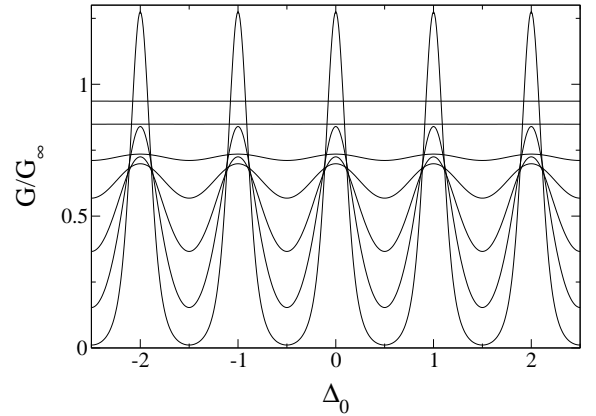}}
\caption{\label{gd0many}
Coulomb blockade oscillations of the conductance as a function
of the dimensionless gate voltage, $\Delta_0$ at $\delta E=E_C$.
Curves are plotted for different temperatures:
$T/E_C=T/\delta E\equiv 1/\delta =0.2, 0.35, 0.5, 0.7, 1, 2, 5$.}
\end{figure}

The linear conductance (\ref{g0}) as a function of the dimensionless 
gate voltage, $\Delta_0$, at $\delta E=E_C$ is plotted in
Fig.~\ref{gd0many} for different temperatures.
At low temperatures there are sharp Coulomb blockade peaks.
At high temperatures, $T\gg E_C,\delta E$, Coulomb blockade is 
lifted and small oscillations of the conductance can be observed. 
These oscillations are slightly non-sinusoidal and given 
by the following asymptotic formula:
\begin{eqnarray}
\frac{G(\Delta_0)-\overline{G(\Delta_0)}}{G_{\infty}}=
2\pi^{3/2}\frac{e^{-\varepsilon_0 /4}}{\sqrt{\varepsilon_0}}
\left[
e^{-\pi^2/\varepsilon_0}\cos (2\pi\Delta_0)
\right.
\nonumber \\
+\left.
e^{-2\pi^2/\varepsilon_0}\cos (4\pi\Delta_0)
+O(e^{-3\pi^2/\varepsilon_0})
\right],
\label{hightexp}
\end{eqnarray}
where $\overline{G(\Delta_0)}$ is the average value of the conductance.
The second term in Eq.~(\ref{hightexp}) is due to inherently non-sinusoidal 
nature of the conductance oscillations, see Fig.~\ref{gd0many}.
To derive this expression one can use Poisson's summation formula.
\begin{figure}[b]
 \resizebox{.42\textwidth}{!}{\includegraphics{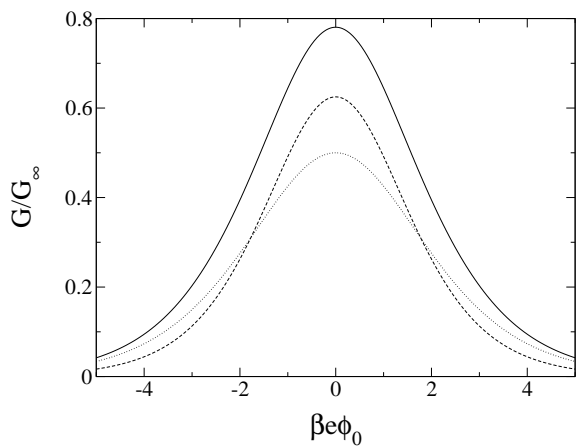}}
\caption{\label{gd0peak}
The exact line shape of the conductance peak, Eq.~(\ref{onepeak}),
plotted at $1/\delta\equiv T/\delta E=0.4$ (solid curve). 
Dotted curve corresponds to blindly applying classical regime 
formula (\ref{classical}) at $T/\delta E=0.4$.
Dashed curve corresponds to blindly applying 
formula (\ref{resonant}) at $T/\delta E=0.4$.
The argument of the plot is linear function 
of the gate voltage, $\phi_0$.}
\end{figure}

To study the line shape of a separate peak let us consider
the limit of large charging energy: $E_C\gg T,\delta E$
or, equivalently, $\varepsilon_0\gg 1,\delta$ in the
dimensionless units. In Eq.~(\ref{g0}) only $N=-1,0$ terms 
in the sum over $N$ give substantial contribution to the
conductance near $\Delta_0=0$; all other terms are exponentially suppressed.
Besides, sum over $j$ at the $N=-1$ is $O(e^{-2\varepsilon_0})$
and, therefore, can also be neglected.
Hence, line shape of the conductance peak at $\Delta_0=0$ is given by
\begin{equation}
\frac{G(\Delta_0)}{G_{\infty}}=
\frac{\delta}{1+e^{-2\varepsilon_0\Delta_0}}
\sum_j\frac{n_j}{1+e^{-j\delta +2\varepsilon_0\Delta_0}}.
\end{equation}
It is more instructive to rewrite this equation as follows:
\begin{equation}
\frac{G(\phi_0)}{G_{\infty}}=
\frac{\delta}{1+e^{-\beta e\phi_0}}
\sum_j\frac{n_j}{1+e^{-j\delta +\beta e\phi_0}},
\label{onepeak}
\end{equation}
where $\phi_0 =V_{e}-V_{e}^{(0)}$; $V_{e}^{(0)}$ is chosen so
that $\phi_0 =0$ corresponds to center of the conductance peak.
In the classical regime, $T\gg\delta E$, line shape of the 
conductance peak is given by~\cite{kulik}
\begin{equation}
\frac{G(\phi_0)}{G_{\infty}}=
\frac{\beta e\phi_0}{2\sinh (\beta e\phi_0)}.
\label{classical}
\end{equation}
In the opposite limit of $\delta E\gg T$:
\begin{equation}
\frac{G(\phi_0)}{G_{\infty}}=
\frac{\delta}{2[1+\cosh (\beta e\phi_0)]}.
\label{resonant}
\end{equation}
The exact line shape of the conductance peak, Eq.~(\ref{onepeak}),
at $T/\delta E=0.4$ is shown in Fig.~\ref{gd0peak}.
On the same figure we also plotted two conductance peaks
in the limiting cases, Eqs. (\ref{classical}) and (\ref{resonant}),
out of their validity region at $T/\delta E=0.4$.
Nevertheless, it is interesting that the exact conductance 
peak is higher than both of the limiting cases peaks.
The peak's height is given by
\begin{equation}
\frac{G(0)}{G_{\infty}}
=\frac{\delta}{2}
\sum_j\frac{n_j}{1+e^{-j\delta}}=
\left\{
\begin{array}{l}
1/2,~~T\gg\delta E \\
\delta /4,~~\delta E\gg T\gg\Gamma_i
\end{array}
\right. .
\label{eqheight}
\end{equation}
Temperature dependence of the conductance peak's height
was numerically calculated in the Ref.~\onlinecite{beenakker}, 
see Fig.~2 there.

\section{Application to tunneling through Quantum
Hall edge states in a quantum dot}

Formulas for the linear conductance in the case of equidistant
energy levels in a dot and spinless fermions 
derived in the previous section can be applied to 
a number of physical problems.

\begin{figure}[b]
 \resizebox{.43\textwidth}{!}{\includegraphics{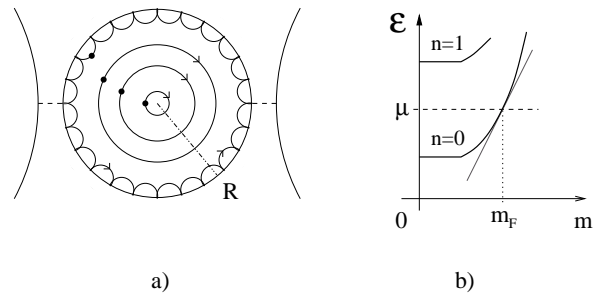}}
\caption{\label{qhe}
Geometry and spectrum of the quantum dot states:
a) symmetric gauge eigenstates, $\left| m\right>$, including 
edge state are shown schematically; b) energy spectrum of the
eigenstates with different angular momenta,
two lowest Landau levels are shown.}
\end{figure}

Let us consider, for example, a quantum dot formed by confining
a two-dimensional electron gas by a circularly symmetric
electrostatic potential, $U(r)$. We assume that $U(r)$ 
is zero at the origin and takes large value at $r=R$, 
where $R$ is the radius of the dot, Fig.~\ref{qhe}a. 
Let us apply a strong magnetic field, $B$, perpendicular 
to the plane of the dot. This situation corresponds to the 
quantum Hall regime and was reviewed in Ref.~\onlinecite{macdonald}.

To solve the one-electron Schr\"odinger equation in this geometry
it is convenient to choose the symmetric gauge. Then, angular
momentum is, clearly, an integral of motion. In each Landau level, $n$,
states with larger angular momentum, $\left| m\right>$, are 
localized further from the origin, near a circle with 
radius $R_m=l_H\sqrt{2(m+1)}$, where $l_H=\sqrt{\hbar c/eB}$ 
is the magnetic length, $c$ is the speed of light. 
The presence of the confinement potential 
leads to an increase in energy for the symmetric gauge 
eigenstates with $R_m$ of order or larger than $R$
(or, $m$ of order or larger than $m_F$, see Fig.~\ref{qhe}b). 
For the states with $R_m\sim R$ an electron is influenced by
both the electric field of the boundary, $E(R)=U'(R)/e$, and 
strong, perpendicular to the electric, magnetic field. 
Thus, near the edge electron executes rapid cyclotron 
orbits centered on a point that slowly drifts in the direction 
of ${\bf E}\times{\bf B}$, that is, along the boundary. 
Thus, Quantum Hall edge states are formed, Fig.~\ref{qhe}a. 
It is important to notice that in this closed geometry
electron system has only one edge.
In this consideration we also assume that $l_H\ll R$.

For simplicity let us consider the case when only the zeroth
Landau level crosses the chemical potential, that is,
there is only one type of edge states. This corresponds
to a sufficiently strong magnetic field so that filling 
factor, $\nu$ is equal to $1$, Fig.~\ref{qhe}b.

Now, we are ready to consider transport through this
type of quantum dot in the strong magnetic field.
Let us weakly couple it to two leads and apply
an infinitesimally small bias voltage between them. 
An electron from the left lead can now tunnel into dot's 
edge state and then tunnel into the right lead as
illustrated by dashed lines in Fig.~\ref{qhe}a.

The energy spectrum of edge states 
can be linearized as follows
\begin{eqnarray}
{\cal E}_{m}=\mu +\left. \frac{\partial {\cal E}}{\partial m}
\right|_{m_F} \left( m-m_F\right).
\end{eqnarray}
Thus, energy levels of the edge states are equally spaced
with the spacing~\cite{macdonald}
\begin{eqnarray}
\delta E=
\left.\frac{\partial {\cal E}_{m}}{\partial m}\right|_{m_F}=
\left.\frac{\partial {\cal E}_{m}}{\partial R_m}\right|_{R}
\left.\frac{\partial R_m}{\partial m}\right|_{m_F}=
eE(R) \frac{l_H^2}{R}.
\label{qhespacing}
\end{eqnarray}
This fact makes formulas derived in the previous section
applicable to this problem. Essential assumption here
is that dispersion curve, Fig.~\ref{qhe}b is almost
linear in the range of angular momentums:
$|m-m_{F}|\lesssim\Delta m$, 
where $\Delta m=\mbox{max}(1,T/\delta E)$.

In the case at hand, spacing, $\delta E$ is inversely 
proportional to the size of a dot just like charging
energy, $E_C$. Hence, their ratio does not depend on
the size of a dot and is given by
\begin{eqnarray}
\frac{\delta E}{E_C}\sim\frac{\epsilon}{\alpha}\frac{E(R)}{B},
\end{eqnarray}
where $\epsilon$ is the dielectric constant of the media 
around the interface;
$\alpha =e^{2}/\hbar c$ is the fine structure constant.
Therefore, in this case oscillations of the conductance
given by Eq.~(\ref{g0}) are determined by only one
parameter $\delta =\delta E/T$.

In conclusion, let us consider the case
of an arbitrary shaped quantum dot.
In this case, $m$ is just the index of an edge state
and no longer associated with the angular momentum. 
The phase along the boundary 
for the $m$-th edge state is
\begin{eqnarray}
\theta_{m}=\int_{0}^{L}dx~k_{m}(x),
\end{eqnarray}
where $k_{m}(x)$ is the corresponding wave vector,
$x$ parametrizes the boundary, and $L$ is its length.
The phase difference between two consecutive edge states is
\begin{eqnarray}
\theta_{m+1}-\theta_{m}=2\pi
=\int_{0}^{L}dx\left[k_{m+1}(x)-k_{m}(x)\right],
\end{eqnarray}
where
$k_{m+1}(x)-k_{m}(x)=\left({\cal E}_{m+1}-{\cal E}_{m}\right)/\hbar v(x)$,
$v(x)=cE(x)/B$ is a drift speed along the boundary.
Then, the spacing between edge states' energy levels is
\begin{eqnarray}
\delta E=2\pi\hbar
\left[\int_{0}^{L}\frac{dx}{v(x)}\right]^{-1}
=2\pi el_H^2\left[\int_{0}^{L}\frac{dx}{E(x)}\right]^{-1}.
\label{qhespacingrealdot}
\end{eqnarray}
Though the electric field $E(x)$ at the boundary
slightly changes as one goes from one edge state to the other,
this effect is small and we neglect it.
Therefore, the energy levels of the edge states 
are equidistant with the spacing given by
Eq.~(\ref{qhespacingrealdot}).

In the case of the circularly symmetric quantum dot,
$E(x)$ is constant, and one can easily perform
the integration in Eq.~(\ref{qhespacingrealdot}).
This leads to the previously obtained expression
for the level spacing, Eq.~(\ref{qhespacing}).

\section{Linear conductance in the spin-$\frac{1}{2}$ case}

Formula (\ref{spinlessg}) for the linear conductance 
in the spinless case can be easily generalized to the 
spin-$\frac{1}{2}$ case by counting each energy level 
twice:~\cite{beenakker}
\begin{eqnarray}
G &=& 2 \frac{e^2}{hT} \sum_{i=1}^{\infty}
\frac{\Gamma_i^l \Gamma_i^r}{\Gamma_i^l + \Gamma_i^r}
\sum_{N_e =1}^{\infty}P_{eq}(N_e)F_{eq}(E_{i\uparrow}|N_e)
\nonumber \\
&& \times [1 - n_F(E_i - \mu + U(N) - U(N-1))],
\label{beespin}
\end{eqnarray}
where $F_{eq}(E_{i\uparrow}|N_e)$ is the occupation number 
of the quantum dot's energy level $i$ with a spin-up electron, 
($i,\uparrow$) in the canonical ensemble: 
number of electrons in the dot, $N_e$, is fixed.

As in the spinless case, to carry out analytical consideration 
we have to assume that energy levels in the quantum dot are 
equally spaced. Presence of the spin degeneracy makes the 
calculations more complicated.

First of all, let us find the occupation number 
$F_{eq}(E_{i\uparrow}|N_e)$.
Let us consider spin-up and spin-down electron subsystems. 
Ground state energy of the system is
$$
E_{g}=\left(N^{2}_{\uparrow}+N^{2}_{\downarrow}\right)
\frac{\delta E}{2}
=\left[
\left(N_{\uparrow}+N_{\downarrow}\right)^{2}
+\left(N_{\uparrow}-N_{\downarrow}\right)^{2}
\right]
\frac{\delta E}{4},
$$
where $N_{\sigma}=(N_e)_{\sigma}-N_{i}/2$ is the number 
of excess electrons in the spin-$\sigma$ subsystem; 
$N_{i}$ is chosen even.
$N_{\uparrow}+N_{\downarrow}\equiv N$ is the total 
number of excess electrons in the quantum dot; 
$(N_{\uparrow}-N_{\downarrow})/2\equiv S_z$
is $z$-component of the total electron spin.
Using these identities, one can find that
\begin{eqnarray}
E_{g}=\frac{1}{4}N^{2}\delta E
+S_{z}^{2}\delta E.
\end{eqnarray}
While $S_{z}$ is subjected to the thermodynamic
fluctuations, $N$ is fixed.
\begin{figure}[b]
\resizebox{.42\textwidth}{!}{\includegraphics{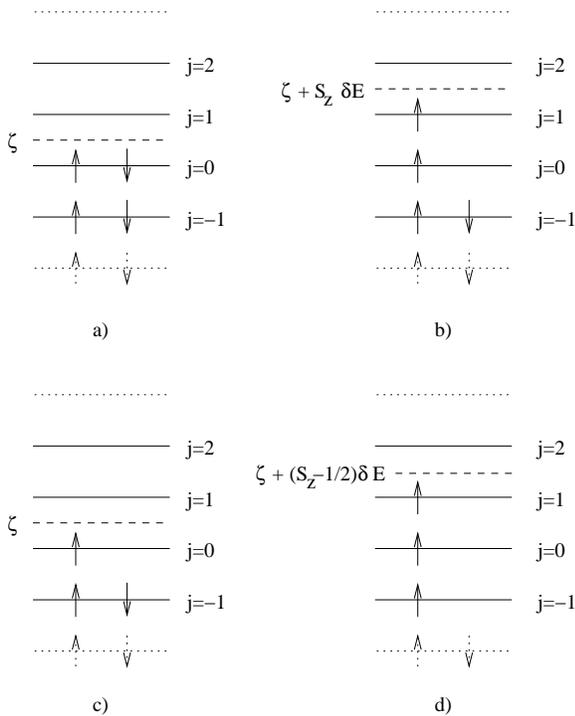}}
\caption{\label{Fig4}
Parameter $\zeta'$ of the spin-up electron subsystem 
for even number of electrons: a) at $S_z = 0$, 
b) at arbitrary integer $S_z$ ($S_z=1$ case is shown); 
for odd number of electrons: c) at $S_z = 1/2$, 
d) at arbitrary half-integer $S_z$ ($S_z=3/2$ case is shown).}
\end{figure}

The occupation number in question, 
$F_{eq}(E_{i\uparrow}|N_e)\equiv n_{j\uparrow}$,
is known if, in addition to $N$, 
the $z$-component of the total spin, $S_{z}$, is fixed. 
In the case of an even number of electrons, 
parameter $\zeta'$ for the spin-up electron subsystem 
is equal to $\zeta +S_{z}\delta E$,
given $S_{z}$, see Figs.~\ref{Fig4}a and \ref{Fig4}b. 
Occupation numbers in the spin-up subsystem at fixed $S_{z}$ 
are given by Eq. (\ref{dent}) 
with the appropriately chosen parameter $\zeta'$: $n_{j-S_{z}}$. 

However, $z$-component of the total spin, $S_{z}$, 
is not fixed but subjected to the thermodynamic fluctuations. 
Therefore, to find the occupation numbers, 
$n_{j\uparrow}^{ev}$, we have to account for all
possible values of $S_{z}$:
\begin{eqnarray}
n_{j\uparrow}^{ev}
=\sum_{S_z=-N_e/2}^{N_e/2}P(S_{z})~n_{j-S_{z}},
\label{nevmod}
\end{eqnarray}
where
\begin{eqnarray}
P(S_{z})=\frac{e^{-S_{z}^{2}\delta}}
{\sum\limits_{S_z=-N_e/2}^{N_e/2}e^{-S_{z}^{2}\delta}}
\label{pszmod}
\end{eqnarray}
is the probability that $z$-component of the total
spin of the quantum dot is equal to $S_{z}$.
Substituting Eqs. (\ref{pszmod}) and (\ref{dent})
in Eq. (\ref{nevmod}) we obtain
$$
n_{j\uparrow}^{ev}
=\sum\limits_{m=1}^{\infty}(-1)^{m-1}
e^{-\left[ m^{2}+(2j-1)m\right]\frac{\delta}{2}}~
\frac{\sum\limits_{S_z=-N_e/2}^{N_e/2}e^{-\left(S_{z}^{2}-mS_{z}\right)\delta}}
{\sum\limits_{S_z=-N_e/2}^{N_e/2}e^{-S_{z}^{2}\delta}}.
$$
Since number of electrons in the quantum dot is even,
$S_{z}$ may take only integer values. Therefore,
$$
n_{j \uparrow}^{ev}=\sum\limits_{m=1}^{\infty}(-1)^{m-1}
e^{-\left[ \frac{m^2}{4}+(j-\frac{1}{2})m\right] \delta}~
\frac{\sum\limits_{S_z = -\infty}^{\infty} 
e^{-(S_z -\frac{m}{2})^2\delta}}
{\sum\limits_{S_z = -\infty}^{\infty} e^{-S_z^2 \delta}},
$$
where we extended limits of the sum over $S_{z}$ to
infinities since the Fermi energy is the largest energy
scale in the problem.
Separating $m=2r-1$ and $m=2r$ parts of the sum, 
where $r$ is a positive integer, we obtain final expression 
for the occupation numbers in the case of even number of electrons:
\begin{equation}
n_{j \uparrow}^{ev} 
= A (\delta) \sum_{r=1}^{\infty} 
e^{-(r-\frac{1}{2})(r+2j-\frac{3}{2})\delta} -
\sum_{r=1}^{\infty} e^{-r(r+2j-1)\delta},
\label{nevanalytic}
\end{equation}
where
\begin{eqnarray}
A(\delta )
&=& \frac{\sum\limits_{s=-\infty}^{\infty} 
e^{-\left( s-\frac{1}{2}\right)^2 \delta}}
{\sum\limits_{s=-\infty}^{\infty} e^{- s^2 \delta}}.
\end{eqnarray}
In two limiting cases
\begin{eqnarray}
A(\delta )
&=& \left\{
\begin{array}{l}
2e^{-\delta /4}
\left[ 1+O\left( e^{-\delta}\right)\right],~~\delta E\gg T  \\
\\
1-4e^{-\pi^2/\delta}+O\left( e^{-2\pi^2/\delta }\right),~~T\gg\delta E
\nonumber
\end{array}
\right. .
\end{eqnarray}
Analytical expression 
for the high-temperature limit of $A(\delta )$ 
can be obtained using Poisson's summation formula.

One can easily prove the following properties 
of the occupation numbers $n_{j \uparrow}^{ev}$
valid at arbitrary temperature:
\begin{eqnarray}
n_{j \uparrow}^{ev} &=& 1 - n_{1-j,\uparrow}^{ev},
\\
e^{j \delta} n_{j \uparrow}^{ev} + e^{-j \delta} n_{-j \uparrow}^{ev}
&=& A( \delta ) e^{\delta /4}.
\end{eqnarray}
They are valid due to the electron-hole symmetry 
and similar to 
the following properties of the Fermi-Dirac distribution:
$n_F(E)=1-n_F(-E)$ 
and $e^{\beta E}n_F(E)+e^{-\beta E}n_F(-E)=1$.

Similarly, one can find occupation numbers 
in the case of odd number of electrons in the quantum dot, $N_{e}$.
Energy level which contains one electron at $T = 0$ 
will be referred to as $j=0$ level. 
In this case electron-hole symmetry corresponds to $j \to -j$ transformation. 
Parameter $\zeta'$ of the spin-up electron subsystem 
at a given $S_z$ is equal to $\zeta +(S_z -1/2)\delta E$, 
see Figs.~\ref{Fig4}c and \ref{Fig4}d.
Therefore,
\begin{eqnarray}
n_{j \uparrow}^{od}
=\sum\limits_{S_{z}=-N_{e}/2}^{N_{e}/2}
P\left( S_{z}\right)~n_{j-\left( S_{z}-\frac{1}{2}\right)}
\nonumber \\
=\sum\limits_{m=1}^{\infty} (-1)^{m-1} 
e^{-\left( \frac{m^2}{4}+jm\right) \delta}
~\frac{\sum\limits_{S_{z}=-N_{e}/2}^{N_{e}/2} 
e^{-\left( S_z -\frac{m}{2}\right)^2 \delta}}
{\sum\limits_{S_{z}=-N_{e}/2}^{N_{e}/2} e^{-S_z^2 \delta}}.
\end{eqnarray}
Since number of electrons in the quantum dot is odd,
$S_{z}$ may take only half-integer values.
Separating odd and even parts of the sum over $m$, we obtain:
\begin{equation}
n_{j \uparrow}^{od}=\frac{1}{A( \delta )}\sum_{r=1}^{\infty}
e^{-\left( r-\frac{1}{2}\right)\left( r+2j-\frac{1}{2}\right)\delta}-
\sum_{r=1}^{\infty} e^{-r(r+2j) \delta}.
\label{nodanalytic}
\end{equation}
Property of the electron-hole symmetry reads as follows:
$n_{j\uparrow}^{od}=1-n_{-j\uparrow}^{od}$.

It turns out that there exists simple relation between 
$n^{ev}$ and $n^{od}$ occupation numbers:
\begin{equation}
n_{j \uparrow}^{ev} =A(\delta ) 
e^{-\left( j-\frac{1}{4}\right)\delta} n_{-j \uparrow}^{od}.
\label{evod}
\end{equation}
This property is the analog of $n_F(E)=e^{-\beta E}n_F(-E)$ 
one of the Fermi-Dirac distribution. It will allow us to get rid of
$n^{od}$ occupation numbers in the final expression for the conductance.

Now we are in a position to find the probability that a dot, 
in thermodynamic equilibrium with the reservoirs, 
contains $N_e$ electrons, $P_{eq}(N_e)$.
Equation (\ref{peq2}) written for the spinless case is still 
applicable if we keep in mind that energy levels in the 
quantum dot are doubly degenerate. Partition function
of the dot's internal degrees of freedom in the canonical
ensemble is
\begin{eqnarray}
{\cal Z}(N_e) &=& \sum\limits_{\{ n_{i\sigma}\}}
e^{- \beta \sum (E_i - \mu)n_{i\sigma}} \delta_{N_e, \sum n_{i\sigma}}
\nonumber \\
&=& \sum\limits_{N_e^{\uparrow}=0}^{N_e} 
\sum\limits_{N_e^{\downarrow}=0}^{N_e}
Z(N_e^{\uparrow}) Z(N_e^{\downarrow})
\delta_{N_e, N_e^{\uparrow}+N_{e}^{\downarrow}},
\label{eq9}
\end{eqnarray}
where
\begin{eqnarray}
Z(N_e^{\uparrow})=\sum_{\{n_i\}}e^{-\beta\sum (E_i - \mu)n_i}
\delta_{N_e^{\uparrow}, \sum n_i}
\end{eqnarray}
is the partition function of the spin-up electron subsystem in 
the canonical ensemble. Mathematically, expression for 
$Z(N_e^{\uparrow})$ is identical to the one for $Z(N_e)$ 
in the spinless case. 
Thus, one can directly apply the result
obtained previously, Eq.~(\ref{zne}):
\begin{eqnarray}
Z(N_e^{\uparrow})=e^{-(N_e^{\uparrow}-N_{i}/2)^2\delta /2}Z_{exc},
\end{eqnarray}
where $N_e^{\uparrow}=N_{i}/2$ is the equilibrium number of 
electrons in the spin-up subsystem. Similar result is valid 
for the partition function of the spin-down electron subsystem, 
$Z(N_e^{\downarrow})$. Substituting these results in Eq.~(\ref{eq9})
we obtain:
\begin{eqnarray}
{\cal Z}(N_e)=Z_{exc}^2 
\nonumber
\\
\times \sum_{N_e^{\uparrow}=0}^{N_e} 
\sum_{N_e^{\downarrow}=0}^{N_e}
e^{-\left[ (N_e^{\uparrow} - N_{i}/2)^2 + 
(N_e^{\downarrow} - N_{i}/2)^2\right]
\delta /2} \delta_{N_e^{\uparrow} + N_e^{\downarrow}, N_e}
\nonumber .
\end{eqnarray}
The exponent can be simplified as follows
$$
\left( N_e^{\uparrow}-\frac{N_{i}}{2}\right)^2 
+\left( N_e^{\downarrow}-\frac{N_{i}}{2}\right)^2
=\frac{N^2}{2}+\frac{\left( N_e^{\uparrow} - 
N_e^{\downarrow}\right)^2}{2},
$$
hence,
\begin{eqnarray}
\frac{{\cal Z}(N_e)}{Z_{exc}^2} = e^{-N^2\delta /4}
\sum_{N_e^{\uparrow}=0}^{N_e}\sum_{N_e^{\downarrow}=0}^{N_e}
e^{-(N_e^{\uparrow}-N_e^{\downarrow})^2\delta /4} 
\delta_{N_e^{\uparrow}+N_e^{\downarrow}, N_e}
\nonumber \\
= e^{-N^2 \delta /4}\sum_{N_e^{\uparrow}=0}^{N_e}
e^{- (N_e^{\uparrow}-N_e /2)^2\delta}
= e^{-N^2 \delta /4} \sum_{s = - N_e /2}^{N_e /2}
e^{-s^2 \delta},
\nonumber
\end{eqnarray}
where in the second equality we took advantage of the delta 
symbol. Sum in the last line is taken over integer values
of $s$ if $N_e$ is even or half-integer values of $s$ if 
$N_e$ is odd. Limits of the sum over $s$ can be
extended to infinities since we assume that $\mu\gg T$.
According to Eq.~(\ref{peq2}) probability that quantum dot, 
in thermodynamic equilibrium with the reservoirs, contains 
$N_e$ electrons is
\begin{eqnarray}
P_{eq}(N_e)=\frac{{\cal Z}(N_e)}{Z_{\mu}}e^{-\beta U(N)}
=\frac{Z_{exc}^2}{Z_{\mu}}
e^{-\beta {\tilde U}(N)}
\sum_{s=-N_e/2}^{N_e/2}e^{-s^2\delta}
\nonumber
\\
=\frac{e^{-\beta {\tilde U}(N)}
\sum\limits_{s=-N_e/2}^{N_e/2}e^{-s^2\delta}}
{\sum\limits_{N=ev}
e^{-\beta {\tilde U}(N)}
\sum\limits_{s} e^{-s^2 \delta}
+\sum\limits_{N=od}
e^{-\beta {\tilde U}(N)} 
\sum\limits_{s}
e^{-\left( s-\frac{1}{2}\right)^2\delta}},
\nonumber
\end{eqnarray}
where ${\tilde U}(N)\equiv U(N)+N^2\delta E/4$;
and we used the fact that $N$ and $N_e$ have the same parity 
since $N_{i}$ is chosen even. 
Therefore, sums over $N=ev$ and $N=od$ are taken over 
$N=0,\pm 2,\pm 4,\ldots$ and $N=\pm 1,\pm 3,\ldots$ values, 
respectively. At this point in the calculation we need to 
specify whether the total number of electrons in the dot 
is even or odd:
\begin{equation}
P_{eq}(N_e)=
\left\{
\begin{array}{l}
\left( D'\right)^{-1}
e^{-\beta {\tilde U}(N)},~~N~\mbox{is~even}
\\
\left( D'\right)^{-1} A(\delta )
e^{-\beta {\tilde U}(N)},~~N~\mbox{is~odd}
\end{array}
\right.
,
\label{pnespin}
\end{equation}
where
\begin{eqnarray}
D'\equiv\sum_{N=even}e^{-\beta {\tilde U}(N)}
+A(\delta )\sum_{N=odd}e^{-\beta {\tilde U}(N)}
\end{eqnarray}

Now we are prepared to calculate the conductance, Eq.~(\ref{beespin}), 
in the case of the equidistant double degenerate energy levels 
in the dot at an arbitrary $\delta E/T$ and
$E_{C}/\delta E$ ratios.
Similarly to the consideration in the spinless case
we assume that quantum dot is weakly coupled to the
leads via multichannel tunnel junctions, and 
tunneling widths of the energy levels in the quantum dot,
$\Gamma_{i}^{l}$ and $\Gamma_{i}^{r}$, are slowly changing
functions of the energy, $E_i$.
\begin{figure}[b]
\resizebox{.48\textwidth}{!}{\includegraphics{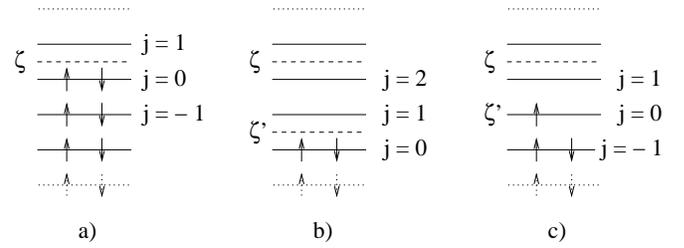}}
\caption{\label{Fig6} 
Mapping of the sum over $i$ onto the sum over $j$ for 
a) $N=0$; 
b) even $N$ ($N=-4$ case is shown); 
c) odd $N$ ($N=-3$ case is shown).}
\end{figure}
Then, these tunneling widths can be considered constants 
and evaluated at the chemical potential:
$\Gamma_i^l\approx\Gamma^l_{\mu}\equiv \Gamma^l$; 
$\Gamma_i^r\approx\Gamma^r_{\mu}\equiv \Gamma^r$.
Furthermore, they can be expressed via
conductances of the corresponding junctions:
\begin{equation}
\Gamma^l=\frac{hG^l}{e^2}\frac{\delta E}{2},
~~~\Gamma^r=\frac{hG^r}{e^2}\frac{\delta E}{2}.
\label{glgr}
\end{equation}
There is an additional factor of $1/2$ here compared to
the spinless case, Eq.~(\ref{glgr0}), due to the double 
degeneracy of each energy level in the quantum dot.
First of all, let us break the sum over $N_e$ in 
Eq.~(\ref{beespin}) in two parts: $N=even$ and 
$N=odd$, and apply Eqs.~(\ref{glgr}) and (\ref{pnespin}):
\begin{widetext}
\begin{eqnarray}
G &=& \frac{G^lG^r}{G^l+G^r}\frac{\delta}{D'}
\left( 
\sum_{N=even}e^{-\beta {\tilde U}(N)}
\sum_{i=1}^{\infty} F_{eq}(E_{i\uparrow}|N_e)
\left\{ 1-n_F\left[ E_i-\mu +U(N)-U(N-1)\right]\right\}
\right.
\nonumber \\
&&
\left.
+A(\delta )\sum_{N=odd}e^{- \beta {\tilde U}(N)}
\sum_{i=1}^{\infty}F_{eq}(E_{i\uparrow}|N_e)
\left\{ 1-n_F\left[ E_i-\mu +U(N)-U(N-1)\right]\right\}
\right).
\label{beforefinal}
\end{eqnarray}
\end{widetext}
To take advantage of the occupation numbers we derived, 
Eqs. (\ref{nevanalytic}) and (\ref{nodanalytic}),
we need to map each of the sums over $i$ onto the sum 
over $j$. The first sum over $i$ in Eq. (\ref{beforefinal}) 
is taken at even number of excess electrons, 
see Figs.~\ref{Fig6}a and \ref{Fig6}b, hence
\begin{eqnarray}
F_{eq}\left( E_{i\uparrow}|N_e\right) &=&n_{j\uparrow}^{ev},
\nonumber \\
E_i-\mu =E_i-\zeta' +\left( \zeta' -\zeta\right) &=&
\left( j-\frac{1}{2}\right)\delta E+\frac{1}{2}N\delta E.
\nonumber
\end{eqnarray}
Remember that by properly choosing ``zero'' 
of the gate voltage we put $\zeta =\mu$.
The second sum over $i$ is taken at odd number of excess electrons,
see Figs.~\ref{Fig6}a and \ref{Fig6}c, therefore
\begin{eqnarray}
F_{eq}\left( E_{i\uparrow}|N_e\right) &=& n_{j\uparrow}^{od},
\nonumber \\
E_i-\mu =E_i -\zeta' +\left(\zeta' -\zeta\right) &=&
j\delta E +\frac{1}{2}N\delta E.
\nonumber
\end{eqnarray}
We will also use the following identities:
\begin{eqnarray}
{\tilde U}(N)&=&\left( E_C +\frac{\delta E}{4}\right) N^2 -eV_{e}N 
\nonumber \\
&=& \left( E_C +\frac{\delta E}{4}\right)
\left( N -\Delta +\frac{1}{2}\right)^2+{\cal C}_{2},
\nonumber
\end{eqnarray}
where ${\cal C}_{2}=-(eV_{e})^{2}/(4E_C+\delta E)$;
\begin{eqnarray}
\Delta\equiv\frac{eV_{e}}{2(E_{C}+\delta E/4)}+\frac{1}{2}
\end{eqnarray}
is the dimensionless gate voltage,
$\Delta =0$ corresponds to a position of the conductance peak; and
\begin{eqnarray}
&& U(N)-U(N-1)=(2N-1)E_C-eV_{e}
\nonumber \\
&& =(2N-1)E_C
-2\left( E_C+\frac{\delta E}{4}\right)\left(\Delta -\frac{1}{2}\right).
\nonumber
\end{eqnarray}
Substituting these results in Eq.~(\ref{beforefinal}) we obtain
\begin{widetext}
\begin{eqnarray}
\frac{G(\Delta )}{G_{\infty}}=
\frac{\delta}{D}
\sum_{N=even} 
e^{-\varepsilon\left( N-\Delta +1/2\right)^2}
\sum_{j=-\infty}^{\infty}n_{j\uparrow}^{ev}
\left[
\frac{1}{e^{-(j-1/4)\delta-2(N-\Delta)\varepsilon}+1}
+\frac{1}{e^{-(j-1/4)\delta+2(N+1-\Delta)\varepsilon}+1}
\right]
,
\label{spinhalfcond}
\end{eqnarray}
where
\begin{eqnarray}
D = \exp \left(\beta{\cal C}_{2}\right) D' 
= \sum_{N=even} e^{-\varepsilon (N-\Delta +1/2)^2}
+ A(\delta ) \sum_{N=odd} e^{-\varepsilon (N-\Delta +1/2)^2};
\end{eqnarray}
\end{widetext}
\begin{eqnarray}
\varepsilon\equiv\beta\left( E_C+\frac{\delta E}{4}\right);
\end{eqnarray}
$G_{\infty}\equiv G^lG^r/(G^l+G^r)$
is the high-temperature, $E_C, \delta E\ll T\ll\mu$, 
limit of the conductance. 
To eliminate $n^{od}$ from the final expression 
we used useful property of the occupation numbers 
given by Eq.~(\ref{evod}).

\begin{figure}[b]
\resizebox{.42\textwidth}{!}{\includegraphics{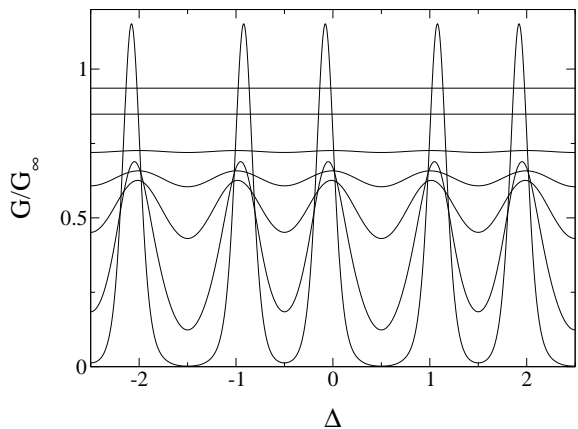}}
\caption{\label{spinhalfplot}
Linear conductance oscillations as a function of the 
dimensionless gate voltage, $\Delta$, at $\delta E=E_C$.
Curves are plotted for different temperatures:
$T/E_C=T/\delta E\equiv 1/\delta =0.15, 0.3, 0.5, 0.7, 1, 2, 5$
using Eq.~(\ref{spinhalfcond}).}
\end{figure}

Formula (\ref{spinhalfcond}) is the main result of this paper.
It is the analytical expression for the linear conductance in the 
spin-$\frac{1}{2}$ case for equidistant energy levels in the quantum dot. 
One can use formula (\ref{spinhalfcond}) to plot Coulomb
blockade oscillations of the conductance as a function
of the dimensionless gate voltage, $\Delta$,
at arbitrary values of $E_C$, $\delta E$ and $T$.
Particularly, when all energy scales are of the same
order: $E_C \sim \delta E \sim T$, numerical calculation
is a breeze, Fig.~\ref{spinhalfplot}.

One can immediately notice the following properties of the 
linear conductance. First of all, $G(\Delta )=G(\Delta +2M)$, 
where $M$ is an integer. In other words, $(4E_C+\delta E)/e$ 
is the conductance period in the gate voltage units. 
This property reflects symmetry with respect to
adding (removing) two electrons to the quantum dot. 
Secondly, conductance is a symmetric function with respect to 
the center of a valley, $\Delta'=M+1/2$, where $M$ is an integer.
That is, $G(\Delta -\Delta')=G(\Delta'-\Delta)$.
This is a reflection of the electron-hole symmetry.
These properties of the conductance oscillations
are not generic. They are valid due to the assumption 
of equally spaced energy levels in a quantum dot.

At high temperatures, $T\gg E_C, \delta E$, conductance 
peaks overlap and their maximums become almost 
equidistant, Fig.~\ref{spinhalfplot}. As a result, instead 
of separate peaks, the conductance in this limit has 
oscillatory behavior.
\begin{figure}[b]
\resizebox{.42\textwidth}{!}{\includegraphics{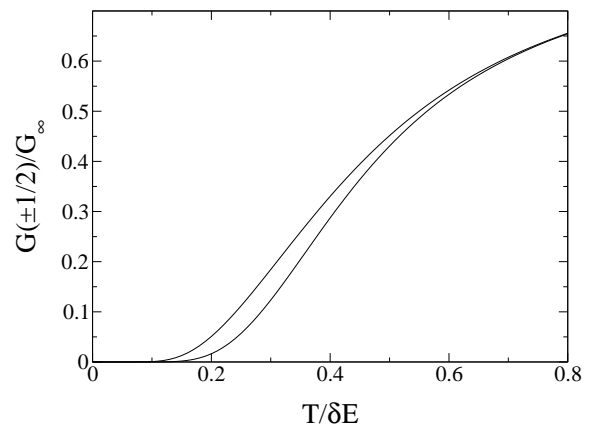}}
\caption{\label{Fig8}
Temperature dependence of the conductance in the odd, 
$G(-1/2)/G_{\infty}$ and even, $G(1/2)/G_{\infty}$ valleys 
at $\delta E=E_C$. Upper curve corresponds to the conductance 
in the odd valley.}
\end{figure}

Let us find temperature dependence of the conductance in 
the valleys. In the sequential tunneling approximation 
the conductance in the valleys decays exponentially as
$T\to 0$, Fig.~\ref{Fig8}. At low temperature number of 
electrons in a dot in the valleys is almost quantized. 
We will call the valley ``odd'' (``even'') if it corresponds 
to odd (even) number of electrons in the dot. We find that 
at any temperature the conductance in the odd valley 
is larger than that in the even one, Fig.~\ref{Fig8}. 
This feature is robust with respect to the distribution of 
energy levels in a quantum dot.

However, it is important to mention that at low temperatures,
$T<T_{in}$, where
\begin{eqnarray}
T_{in}\simeq\frac{E_C}{\ln
\left(
\frac{e^2/\hbar}{G^l+G^r}
\right)}
\end{eqnarray}
cotunneling\cite{glazmannato} will dominate sequential
tunneling contribution to the conductance in the valleys.
Therefore, temperature dependence of the conductance in 
the valleys, Fig.~\ref{Fig8}, is valid only for
the temperatures $T>T_{in}$.

Let us analyze the limit of large charging energy: 
$E_C\gg T, \delta E$ or, equivalently, 
$\varepsilon\gg 1, \delta$ in the dimensionless units.
In this limit, two adjacent peaks in the conductance have 
exponentially small, $\sim e^{-E_C/T}$, overlap with each other.
Thus, it makes perfect sense to study the line shape of a 
separate peak. Let us determine line shape of the conductance
peak near $\Delta =0$. In the numerator of Eq. (\ref{spinhalfcond}) 
only the $N=0$ term in the sum over $N$ survives;
moreover, at $N=0$ second term in square brackets 
is $O(e^{-2\varepsilon})$. 
In the denominator, only the $N=-1,0$ terms matter.
Hence, the line shape of the conductance peak near $\Delta =0$ at 
arbitrary $\delta E/T$ ratio is given by
\begin{equation}
\frac{G(\phi )}{G_{\infty}}=
\frac{\delta}{1+A(\delta )e^{-\beta e\phi}}
\sum_{j=-\infty}^{\infty}
\frac{n_{j\uparrow}^{ev}}{1+e^{-(j-1/4)\delta +\beta e\phi}},
\label{peakspinexact}
\end{equation}
where we used the following identity:
$$
2\varepsilon\Delta 
= \beta e\left( V_{e}-V_{e}^{(0)}\right)
= \beta e\phi .
$$
Clearly, $\phi =0$ corresponds to $\Delta =0$.
In the classical regime, $T\gg \delta E$, the line shape 
of the conductance peak is given by~\cite{kulik}
\begin{equation}
\frac{G(\phi )}{G_{\infty}}=
\frac{\beta e\phi}{2\sinh (\beta e\phi )}.
\label{spinclassical}
\end{equation}
Formally, this equation is identical to that of the spinless case, 
Eq.~(\ref{classical}). Nonetheless, the values of $G_{\infty}$ 
are different in these two cases by a factor of $2$. 
This is due to spin degeneracy of each energy level 
in the spin-$\frac{1}{2}$ case, 
compare Eqs.~(\ref{glgr0}) and (\ref{glgr}). 
In the limit of $\delta E\gg T$:~\cite{glazman}
\begin{equation}
\frac{G(\phi )}{G_{\infty}}=
\frac{\delta}{3+2\sqrt{2}\cosh 
\left(
\beta e\phi +\frac{1}{4}\delta -\frac{1}{2}\ln 2 
\right)}.
\label{eq12}
\end{equation}
\begin{figure}[b]
\resizebox{.42\textwidth}{!}{\includegraphics{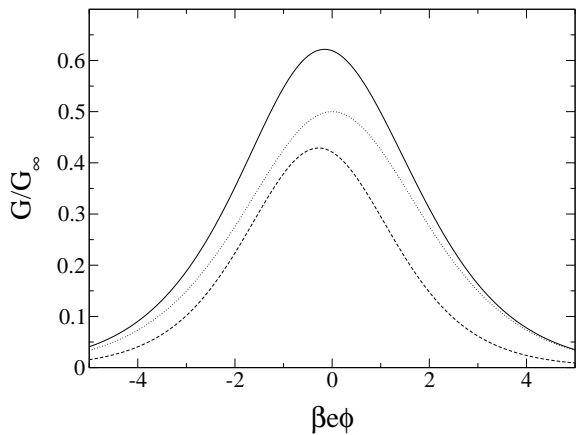}}
\caption{\label{Fig9} 
The exact line shape of the conductance peak, 
Eq.~(\ref{peakspinexact}), plotted at 
$1/\delta\equiv T/\delta E =0.4$ (solid curve).
Dotted curve corresponds to blindly applying 
classical regime formula, Eq.~(\ref{spinclassical}),
at $T/\delta E =0.4$.
Dashed curve corresponds to blindly applying 
formula (\ref{eq12}) at $T/\delta E=0.4$. 
The argument of the plot is linear function of 
the gate voltage, $\phi$.}
\end{figure}
This peak has its maximum at 
\begin{equation}
e\phi_{LT}\equiv
e\phi_{m}\left( T\ll\delta E\right)
=-\frac{1}{4}\delta E+\frac{\ln 2}{2}T,
\label{phirt}
\end{equation}
and is symmetric with respect to this value:
$G(\phi -\phi_{LT})=G(\phi_{LT}-\phi )$.
The exact line shape of the conductance peak, 
Eq.~(\ref{peakspinexact}), at $T/\delta E=0.4$ and
two limiting cases conductance peaks, 
Eqs.~(\ref{spinclassical}) and (\ref{eq12}), plotted
out of their validity region at $T/\delta E =0.4$
are shown in Fig.~\ref{Fig9}.
As in the spinless case, the exact conductance
peak is higher than both of the limiting cases
peaks.

\begin{figure}[b]
\resizebox{.42\textwidth}{!}{\includegraphics{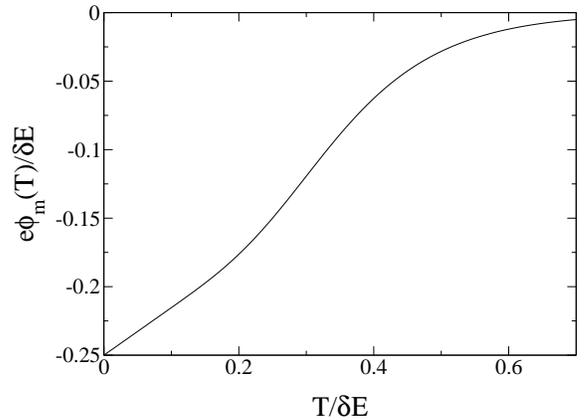}}
\caption{\label{Fig10}
Temperature dependence of the dimensionless
peak's maximum position, $e\phi_m(T)/\delta E$.
In the low temperature limit: $e\phi_m(0)/\delta E=-1/4$
according to Eq. (\ref{phirt}).}
\end{figure}

The position of the peak's maximum, $\phi_m =\phi_m (T)$,
is shifted to the left from its high temperature limit,
$\phi_{CL}\equiv\phi_m(T\gg\delta E)=0$.
It is determined by the equation for $\phi_m$:
$G'(\phi_m )=0$, where $G(\phi )$ is given by 
Eq.~(\ref{peakspinexact}).
The dimensionless position of the peak's maximum, 
$e\phi_m(T)/\delta E$, as a function of the temperature, 
$T/\delta E$, is numerically plotted in Fig.~\ref{Fig10}.

\begin{figure}[b]
\resizebox{.42\textwidth}{!}{\includegraphics{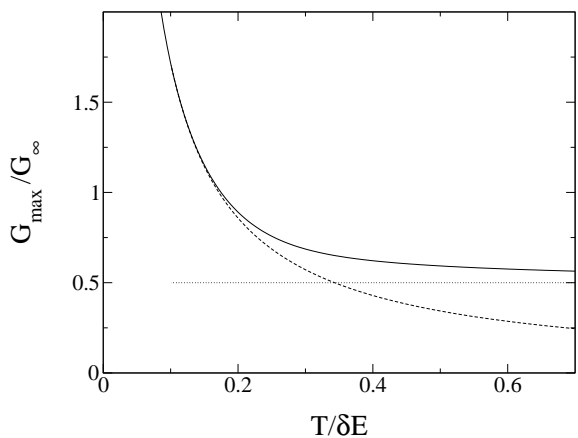}}
\caption{\label{heightspin} 
Height of the conductance peak, $G_{max}/G_{\infty}$, as a 
function of the temperature, $T/\delta E$. 
Dashed curves correspond to two limiting cases peak heights, 
Eq.~(\ref{spheight}).}
\end{figure}

The conductance peak height is $G_{max}=G(\phi_m)$. In the
limiting cases:
\begin{equation}
\frac{G_{max}}{G_{\infty}}=\frac{G(\phi_m)}{G_{\infty}}=
\left\{
\begin{array}{l}
1/2,~~T\gg\delta E \\
(3-2\sqrt{2}) \delta,~~\delta E\gg T\gg\Gamma_i
\end{array}
\right.
.
\label{spheight}
\end{equation}
Peak's height as a function of the temperature, $T/\delta E$,
can be plotted numerically, Fig.~\ref{heightspin}.

\section{Conclusions}

We have studied Coulomb blockade oscillations of the
linear conductance through a quantum dot weakly
coupled to the leads via multichannel tunnel
junctions in the sequential tunneling approximation. 
To obtain analytical results we have assumed that 
the energy levels in the dot are equally spaced.
The electron-electron interaction in a quantum dot 
has been described by the constant interaction model;
though, thermal excitations with all possible spins 
have been taken into account.

The linear conductance in the spinless case is given by Eq.~(\ref{g0}).
It is valid at arbitrary values of $E_C$, $\delta E$ and $T$.
The line shape of an individual conductance peak at arbitrary
ratio $\delta =\delta E/T$ is given by Eq.~(\ref{onepeak}).
Exact conductance peak is higher than both of the 
limiting cases peaks at any gate voltage
as is illustrated in Fig~\ref{gd0peak}.
An analytical expression for the height of the conductance
peak at any ratio $\delta$ is obtained, Eq.~(\ref{eqheight}).

In Sec.~IV we applied the spinless case theory result
to the problem of 
the transport via a dot in the quantum Hall regime.
Energy levels in a dot in this case are equidistant
with the spacing given by Eq.~(\ref{qhespacingrealdot}).

Linear conductance in the case of spin-$\frac{1}{2}$ 
electrons at arbitrary values of $E_C$, $\delta E$, and $T$ 
is given by Eq.~(\ref{spinhalfcond}).
In particular, this equation allows one to plot
the conductance oscillations in the regime when
the charging energy, level spacing in the dot, and
the temperature are all of the same order, Fig.~\ref{spinhalfplot}.
We find that the period of Coulomb blockade
oscillations is doubled compared to the model with
a continuous electronic spectrum in the dot.
Equation~(\ref{spinhalfcond}) is the main result of the paper.

We also find that conductance in the odd valley is larger 
than that in the even one at any temperature, Fig.~\ref{Fig8}.
The difference between conductances has the largest value at
$T\approx 0.3~\delta E$ (at $\delta E=E_C$). 
The sign of the difference is the same as for the quantum dot 
in the Kondo regime.\cite{glazmannato}
Kondo effect takes place at very low temperatures,
$T\lesssim T_{K}\ll T_{in}$,
where $T_K$ is the Kondo temperature,
and leads to the logarithmic enhancement
of the conductance in the odd valley.\cite{aleiner}
Our consideration shows that even-odd asymmetry
exists at much higher temperatures.

Line shape of the conductance peak is given by 
Eq.~(\ref{peakspinexact}). As in the spinless case,
the conductance peak is higher than both of the 
limiting cases peaks at any gate voltage, Fig~\ref{Fig9}.
As we increase the temperature peaks' maximums shift and
become more equidistant, Fig~\ref{Fig10}.
The peak's height as a function of the temperature
is calculated numerically and plotted in Fig.~\ref{heightspin}.

Though we have found physical system which has equidistant energy 
levels in the spinless case, see Section IV, we are not aware of 
any such system in the spin-$\frac{1}{2}$ case. In the case of 
a chaotic quantum dot Wigner-Dyson model gives a fairly good 
approximation for the distribution of the energy levels of the dot. 
If we had assumed Wigner-Dyson distribution of the quantum dot's 
energy levels then we would have had to give up the hope of finding 
a solution. It goes back to the very difficult problem of finding 
occupation numbers of the dot's energy levels in the canonical ensemble. 
The only way to solve it is to assume that energy levels in the quantum dot 
are equally spaced. Then one can use the bosonization technique to find 
the occupation numbers. Assumption of the equidistant energy levels is in 
line with the level repulsion property of the Wigner-Dyson distribution. 
Therefore, the analytical consideration of this reasonably simplified model, 
in our opinion, is a significant step forward in the solution of the 
general problem.

Though we do not expect our quantitative results to precisely describe 
a quantum dot with random energy levels, they certainly give correct order 
of magnitude for the conductance oscillations and their generic features.

\begin{acknowledgments}
The author is greatly indebted to Prof. K.A.~Matveev for drawing
his attention to the problem and most stimulating discussions.
The author is grateful to Prof. A.~Bezryadin for helpful 
and stimulating discussions;
Prof. H.U.~Baranger for valuable comments on the manuscript
and his support; and
Prof. S.W.~Teitsworth for helpful comments on the manuscript.
This work was supported in part by the NSF Grants DMR-9974435
and DMR-0103003.
\end{acknowledgments}

\appendix*
\section{Occupation numbers in the canonical ensemble}

In the canonical ensemble, where the total number of 
particles, $N_e$, is fixed, it is difficult to calculate 
fermionic occupation numbers, Eq.~(\ref{tough}), directly. 
This is true even in the case of equidistant energy levels
separation.

Fortunately, the bosonization technique allows one to express 
fermionic field annihilation and creation operators,
$\psi$ and $\psi^{\dagger}$,
in terms of the bosonic annihilation and creation,
$a_q$'s and $a_q^{\dagger}$'s, and ladder, $U$ and $U^{\dagger}$,
operators:\cite{haldane}
\begin{eqnarray}
\psi^{\dagger}(x) = \frac{1}{\sqrt{L}} e^{-i k_F x}
e^{-i \chi^{\dagger}(x)} U^{\dagger} e^{-i \chi (x)},
\end{eqnarray}
where
\begin{eqnarray}
\chi^{\dagger} (x) = \frac{\pi x}{L} N + i \sum_{q>0}
\sqrt{\frac{2 \pi}{q L}}~e^{-i q x} a_q^{\dagger};
\end{eqnarray}
$L$ is the length of the artificial system; $k_F$ is the Fermi wave vector; 
$N=N_{e}-N_{i}$ is the number of excess electrons operator; 
$q$ is the wave vector; 
and $x$ is the coordinate.
Since these bosons naturally exist in the grand canonical 
ensemble, one can fix total number of excess electrons, $N$,
and calculate occupation  numbers in the bosonic basis,
Eq.~(\ref{dent}).

Occupation numbers, Eq.~(\ref{dent}), have the following 
properties:
\begin{eqnarray}
n_j = 1 - n_{1-j},
~~~~n_j e^{j \delta / 2} = n_{-j} e^{-j \delta / 2}.
\end{eqnarray}
They are similar to $n_F(E)=1-n_F(-E)$ and 
$n_F(E)e^{\beta E/2}=n_F(-E)e^{-\beta E/2}$ 
ones of the Fermi-Dirac distribution and reflect the 
electron-hole symmetry. Combining these two properties 
one can get the recursion relation: 
\begin{eqnarray}
n_{j+1} = 1 - e^{j \delta} n_j.
\end{eqnarray}
In the limit of low temperature, $\delta E \gg T$, for $j>0$
we obtain
\begin{eqnarray}
n_j = e^{- j \delta} - e^{-(2j+1)\delta} 
+O\left[ e^{-(3j+3)\delta}\right];
\end{eqnarray}
in the high temperature limit, $T\gg\delta E$:
\begin{eqnarray}
n_j=\frac{1}{e^{\left( j-\frac{1}{2}\right)\delta}+1} 
-\frac{\delta}{8}
\frac{\sinh\left[\left( j-\frac{1}{2}\right)\frac{\delta}{2}\right]}
{\cosh^3 \left[\left( j-\frac{1}{2}\right)\frac{\delta}{2}\right]}
+O\left(\delta^2\right);
\end{eqnarray}
thus, we find first correction to the Fermi-Dirac distribution.
The form of the correction remains valid even for the slightly
non-equidistant energy levels in the dot. 
If this is the case we need to define ratio $\delta$
in the last expression
as $\delta =\overline{\delta E}/T$, where 
$\overline{\delta E}$ is the mean level spacing.

\end{document}